\documentclass[useAMS,usenatbib]{mn2e}
\usepackage{graphicx}
\usepackage{amssymb}
\usepackage{lscape}
\usepackage{rotating}

\def\deg{\ifmmode^\circ\else$^\circ$\fi}

\def\msun{M$_{\odot}$}

\def\Q{\ifmmode\mathcal{Q}\else$\mathcal{Q}$\fi}
\def\Mach{\ifmmode\mathcal{M}\else$\mathcal{M}$\fi}

\title[{\it Spitzer} IRAC study of AFGL 437]
{{\it Spitzer} IRAC imaging photometric study of the massive star forming region AFGL 437}

\author[L.K. Dewangan \& B.G. Anandarao]
{Lokesh Kumar Dewangan$^{1}$\thanks{lokeshd@prl.res.in}, \& B.G. Anandarao$^{1}$\thanks{anand@prl.res.in}
\\
$^1$Astronomy $\&$ Astrophysics Division, Physical Research Laboratory, Navrangpura, Ahmedabad 380 009, India.\\
}
\begin{document}

\date{ }

\pagerange{\pageref{firstpage}--\pageref{lastpage}} \pubyear{2009}

\maketitle

\label{firstpage}

\begin{abstract}
We present {\it Spitzer} IRAC mid-infrared photometry 
on the massive star forming region AFGL 437 (IRAS 03035+5819). From the IRAC colour-colour diagram, 
we identify several new embedded YSOs within 64 arcsec of the central compact cluster. 
Using the IRAC ratio images, we investigate the molecular outflows associated with the highly embedded 
young stellar object WK34 in the central cluster. 
We attribute the lobes seen (extended to $\sim$ 0.16 pc in the north) in the ratio map 
to shocked molecular hydrogen emission. 
IRAC images reveal a large diffuse nebulosity associated with the central 
cluster and extending up to $\sim$ 8.0 pc from south-west to north-east direction with its  
brightness gradually increasing from 3.6 to 8.0 $\mu$m. A dense box-car-shaped nebula (more than 2.0 pc in size)
situated to the south-west of the cluster shows molecular hydrogen emission 
that may have been caused by shock waves from the compact cluster
sources. It seems that these sources are also responsible for the infrared-bright nebulosity. 
Using a 2D radiative transfer model, we derive from the spectral energy 
distributions, the mass, age and luminosity 
of all the YSOs identified within the central cluster. The SED modelling shows that the 
driving engine of the outflows, WK34, appears to be massive but very young and deeply embedded. 
The weighted mean values of the masses and ages of the 21 YSOs derived from the model are in the range 
1-10 \msun\ and 10$^{4.1 - 6.4}$ yr respectively; while 
their luminosities are in the range of 10$^{0.47-3.48}$ L$_{\odot}$.  
 
\end{abstract}


\section{Introduction}
\label{sec:intro}
Compact clusters embedded in giant molecular clouds provide an opportunity to study 
recent star formation over a wide range of masses in a small volume (e.g., \citet{Lada03}). 
Due to the large visual extinction suffered by the protostars in such clusters, infrared, especially 
mid- and far-infrared, observations help in studying them.  
AFGL 437 (IRAS 03035+5819; G139.909+0.197) is one such compact 
embedded cluster of size $\sim$ 15 arcsec \citep{Kleinmann77, Lada03}
situated at a distance of 2.0 $\pm$ 0.5 kpc \citep{Arquilla84} and having 
a total luminosity of $\sim$2 $\times$ 10$^{4}$ L$_{\odot}$ \citep{Wein96}.
The cluster is associated  with an optical reflection nebula and contains 
at least four embedded sources called AFGL 437N, S, E and W \citep{Cohen77}. 
\cite{Rainer87} and later \citet{Wein96} resolved a highly embedded source known
as WK 34, $\sim$2.5 arcsec to the South-East of AFGL 437N.
The bright sources in the cluster suffer a large extinction of A$_{v} \sim$ 7 mag \citep{Cohen77}.
Radio observations revealed that AFGL 437W and S, identified as 
early B type ZAMS stars, are associated with Ultra-Compact (UC) H II 
regions \citep{Wynn81, Torrelles92, Kurtz94}. 
\citet{Wynn81} and \citet{Torrelles92} have detected water masers towards
AFGL 437N and W, which indicate the youth of the region of star formation.
\cite{Gomez92} found a compact, poorly collimated bipolar CO outflow 
oriented in North-South direction.
Studies of the cluster in near-IR polarimetry \citep{Wein96} 
and diffraction-limited imaging in 3.8 $\mu$m \citep{Weinet96}
have attributed the outflow to the highly-embedded, low-luminosity source WK 34; 
this was later confirmed by the HST polarimetric
imaging study \citep{Meakin05}. HST results showed further that the reflection nebulosity
is due to the source WK 34.
\cite{Davis98} found H$_{2}$ emission `wisps' towards the south-west of the compact cluster, but not     
associated with the outflow. Recent high-resolution C$^{18}$O study of the region
by \cite{Saito06, Saito07} revealed a few dense cores/clumps around the 
central cluster in AFGL 437, indicating  the activity of 
star formation in the region. 
\cite{deWit09} provided spatially resolved 24.5 $\mu$m observation on the main sources of
the cluster. Sub-mm/mm observations of the cluster were made by 
\cite{Dent98} using beam sizes of 16-19 arcsec which 
do not resolve the cluster members. 
\citet{Devine08} determined the age of the cluster to be 1-5 Myr.   
In the background of a number of these important observations and inferences made already on this interesting 
massive star forming region, our motivation for the present study has been to look into the available 
(space) infrared imaging observations beyond 3 $\mu$m, which are not studied so far, 
in order to understand the evolutionary stages of the highly embedded cluster members as well as 
the regions beyond the cluster and compare with the earlier results.  

{\it Spitzer} Infra-red Array Camera (IRAC) provides an opportunity with an 
unprecedented high spatial resolution in thermal infrared wavelength regime that is
very useful in identifying embedded sources in massive star forming regions.
IRAC has four wavelength bands (with $\lambda_{eff}/\Delta\lambda$, 3.55/0.75, 4.49/1.0, 5.73/1.43 and 7.87/2.91 $\mu$m) 
which include molecular emissions such as those from H$_{2}$  
and Polycyclic Aromatic Hydrocarbon (PAH) molecules.
The aims of the present study are to identify the embedded sources using the IRAC bands
in order to classify the different stages of their evolution;
to use ratio maps of the four IRAC bands in order to identify possible H$_{2}$ emission regions 
(or PAH regions) following the suggestion of \cite{Smith05} and \cite{Povich07}; 
and to present the nebulosity associated with the cluster
in all the IRAC bands.
Further, we construct spectral energy distributions (SEDs) of the identified
embedded sources in the cluster using not only the IRAC channels but also observations in JHK bands
\citep{Wein96,Meakin05} and mid-IR \citep{deWit09} and sub-mm/mm \citep{Dent98} regions; and
model the SEDs using the on-line tool developed by \cite{Robit06}.

In Section 2, we describe the data used for the present study 
and the analysis tasks utilised. Section 3 presents the results on 
{\it Spitzer} IRAC photometry of 21 embedded sources associated with the dense
young cluster. In the same section, we discuss our results 
in the light of what is already known about AFGL 437. 
Further in Section 3, we present
the results and discussion on the ratio maps and on the SED modelling of
the YSOs. In Section 4, we give the conclusions.

\section{{\it Spitzer} IRAC Data on AFGL 437 and Data Reduction}
\label{sec:data}
The {\it Spitzer} Space Telescope (SST) IRAC archival
images were obtained from the Spitzer public archive, using `leopard' software (see \citet{Fazio04}, for 
details on the IRAC instrument). The observations relevant for AFGL 437 
were taken in High Dynamic Range (HDR) mode with 12s integration time in all
filters. These observations were a part of the project entitled, ``The Role of Photo-dissociation Regions in High
Mass Star Formation" (Program id 201; PI: G. Fazio).  
Basic Calibrated Data (BCDs) images
were processed for `jailbar' removal, saturation and `muxbleed' correction before making the final mosaic using Mopex
and IDL softwares. A pixel ratio (defined as the ratio of the area formed by the original pixel scale, 1.22
arcsec/px, to that of the mosaiced pixel scale) of 2 was adopted for making the mosaic 
(giving a mosaic pixel scale of 0.86 arcsec/pixel) \footnote[1]{see 
http://ssc.spitzer.caltech.edu/postbcd/doc/mosaiker.pdf}. Using these procedures, a total number of 
60 BCD images of 5.2 $\times$ 5.2 arcmin$^{2}$ were mosaiced to make a final image of 17.4 $\times$ 14.2 arcmin$^{2}$
in each of the four bands. 
Aperture photometry was performed on the mosaic with 2.8 pixel aperture and sky 
annuli of 2.8 and 8.5 pixels using APPHOT task in IRAF package. The zero points for these apertures (including
aperture corrections) were, 17.08, 17.30, 16.70 and 15.88 mag  for the 3.6, 4.5, 5.8, 8.0 $\mu$m bands, 
here onwards called as Ch1, Ch2, Ch3 and Ch4 respectively. The photometric uncertainties 
vary between 0.01 to 0.22 for the four channels, with Ch3 and Ch4 on the higher side.
Ratio maps were produced from the IRAC channel images by using the standard procedure, 
with a pixel ratio of 8 that reveals the features prominently. 

For the purpose of SED modelling, the JHK photometric 
data were taken from \cite{Wein96} and \cite{Meakin05} as well as from 2MASS archives
\citep{Skru06}. 
The 24.5$\mu$m photometric fluxes were obtained from \cite{deWit09}
(at a diffraction-limited resolution of 0.6 arcsec), for the three main sources, WK34, S and W, in the 
compact central cluster. 
In addition, the sub-mm/mm data from \citet{Dent98} were used as upper limits 
on the central sources, being unresolved in the large beam sizes. 
The details of the data compiled on individual sources are given in section 3.3.
We modelled the SEDs so constructed using the on-line SED-fitting tool due of \cite{Robit06}. 
A criterion $\chi^{2}$ - $\chi^{2}_{best}$ $<$ 3 was chosen to obtain weighted means and
standard deviations of individual physical parameters from sets of models for each object. 
For the SED modelling, we followed a similar procedure as described by \cite{Grave09}.

\section{Results and Discussion} 
\label{sec:discussion}
Fig 1 shows the IRAC images of the entire extent of AFGL 437 in all four bands.
The central compact cluster is shown 
marked by a square box in the Ch3 image (bottom left).  
As shown in Fig 1, the IRAC 8.0 $\mu$m band image (bottom right) reveals a diffuse 
bubble-/fan-like nebulous structure associated 
with and extended from the compact cluster in the south-west to 
the north-east direction (with a size of $\sim$ 8.0 pc). 
We find that brightness of the nebulosity 
gradually increases from 3.6 to 8.0 $\mu$m (Ch1-Ch4), as seen in Fig 1. 
One can also notice in Fig 1, several filamentary
structures probably due to the UCHII regions associated with 
AFGL 437S and W in the compact cluster.
Just below the box in the south-west in Fig 1 (between the two arrows in Ch3 image in bottom left), 
one can notice a dense cross-/boxcar-shaped structure that seems to be expanding 
into the surrounding interstellar medium (ISM). It has bright 
components in NE \& SW (separated by $\sim$ 1.70 pc) 
and in SE \& NW (separated by $\sim$ 1.5 pc) directions. 
The corners of this boxcar-like structure  
are seen faintly (as `wisps') in the narrow band H$_{2}$ image 
by \cite{Davis98}. We find that this feature becomes more prominent in longer wavelengths (see Fig 1).
It may be possible that this was generated through interaction of stellar wind from 
the sources AFGL 437W \& S with the surrounding dense material. 
 The compact cluster (inside the box in Fig 1: Ch3) is shown enlarged  
in Fig 2 (in a colour composite of three images: Ch1 (blue), Ch2(green) and Ch4 (red)), 
in which we have marked the central main sources 
namely AFGL 437 N, S, W and WK34 (E is not detected in IRAC) as well 
as other YSOs identified in the region (see Section 3.1). 

The compact cluster appears very nebulous in all the four bands. 
As shown by the earlier authors (e.g., \citet{Wynn81}), 
the source AFGL 437W is associated with a blister HII region. It appears like 
a diffuse source with its brightness increasing progressively in the four IRAC channels Ch1-4
(and is also seen in the 24.5 $\mu$m image by \citet{deWit09}).
A dense filamentary structure, seen to the right of AFGL 437W in all the bands, 
is probably associated with the source and 
extends to about 38.3 arcsec (0.4 pc) in NE-SW direction (see Section 3.2).

\subsection{Cluster Sources}
IRAC [3.6]-[4.5] vs [5.8]-[8.0] colour-colour diagram is shown to be a very powerful tool to classify
proto-stellar objects into their evolutionary stages, such as Class~0/I, Class~II and Class~III \citep{Allen04}. 
Such a colour-colour diagram for the sources is shown in Fig~\ref{fig3}, along with divisions for 
various pre-main-sequence classes, shown by boxes for 
Class~0/I, Class~II and Class~III sources.
The criteria for Class~I/II, which shares the properties of Class~I and Class~II, are taken from \citet{Megeath04}.
Based on the above criteria, we identify 13-14 sources 
(including AFGL 437S and WK34) as Class~0/I and 4-5 as Class~I/II including AFGL 437W, 
in the vicinity (within 90 arcsec) of the central compact cluster (see Fig~\ref{fig2}). 
Table~\ref{tab1} gives the photometric magnitudes of these sources,  
marked as s1, s2 etc. in Fig~\ref{fig2}. It may be  noted that only 18 sources were detected in all the four channels.
We have included in Table 1 three sources (s2, s3 and AFGL 437N) that were not detected in all the four bands, but   
satisfied one of the two colour criteria for Class~I/II or Class~II. In the Object column in Table 1, 
the numbers in parentheses correspond to the sources 
identified earlier by \citet{Wein96} (see their Table 1 and Fig 2). Some of these 
sources are seen in the K band image of \citet{Wein96}, but are not detected in J and H bands. 
From IRAC photometry, we have identified a number of new YSOs, apart from the sources
AFGL 437W, S, WK34, s2, s5-s8 and s10 that are classified earlier by \citet{Wein96}.  
A large number of these sources are highly embedded and suffer large extinction. 
We have determined the visual extinction A$_{v}$ for the sources that have JHK 
photometry \citep{Wein96,Meakin05} and found that WK34 is the most embedded of all,  
with A$_{v}$ = 30-35 mag. 
In addition to the Class I and Class I/II cluster sources, about 35 Class~II
sources (from Fig 3) are also identified outside the cluster, 
spread over the nebular bubble. These are shown as open circles in Ch1 image in Fig 1 (top left). 
Most of these sources occur within a distance of about 1.6 pc (165.4 arcsec) 
southeast of the cluster. For about 23 of these sources,  
good quality 2MASS JHK photometric colours provide confirmation of their low mass Class II nature.
This supports our conjecture that the filamentary structures and the expanding 
boxcar-like nebula are possibly driven by the central cluster sources.

\begin{table*}
\centering
\caption{{\it Spitzer} IRAC 4-channel photometry (in mag) of the YSOs identified in the central cluster of AFGL 437; 
the numbers in parentheses in the Object column are Weintraub \& Kastner (1996) designations (see text)}
\label{tab1}
\begin{tabular}{lccccc||cccccc}
\hline
Object        &   RA [2000]       &   Dec [2000]       & Ch1   &   Err-Ch1 & Ch2     & Err-Ch2 & Ch3   & Err-Ch3 & Ch4     & Err-Ch4 &class   \\

\hline                                                                                                                     
s1 (22)  & 3:07:20.81 & 58:30:35.60 & 12.65 &  0.03  &  11.97  & 0.01   &  11.56  &  0.19  &  10.02  &  0.12    & I      \\
s2 (23)  & 3:07:22.53 & 58:30:45.58 &  9.65 &  0.01  &   8.97  & 0.01   &   7.97  &  0.04  &   ---   &    ---   & ---    \\
s3 (29)  & 3:07:23.04 & 58:30:48.90 &  8.86 &  0.04  &   8.16  & 0.03   & 	---  &   ---  &   ---   &    ---   & ---    \\
W  (28)  & 3:07:23.88 & 58:30:50.25 &  8.34 &  0.05  &   7.77  & 0.05   &   5.50  &  0.05  &   3.52  &  0.05    & I      \\
N  (35)  & 3:07:24.29 & 58:30:54.84 &   --- &   ---  &    ---  &  ---   &   5.23  &  0.04  &   3.34  &  0.04    & ---    \\
S  (20)  & 3:07:24.53 & 58:30:42.90 &  6.72 &  0.01  &   5.46  & 0.01   &   4.11  &  0.00  &   2.68  &  0.01    & I      \\
WK34 (34)  & 3:07:24.55 & 58:30:52.76 &  7.24 &  0.03  &   6.05  & 0.01   &   4.98  &  0.03  &   3.51  &  0.05    & I      \\
s4        & 3:07:25.94 & 58:30:08.90 & 12.36 &  0.02  &  10.55  & 0.01   &   9.26  &  0.01  &   8.35  &  0.04    & I      \\
s5 (7)   & 3:07:26.21 & 58:30:20.96 &  9.71 &  0.02  &   9.54  & 0.02   &   6.52  &  0.02  &   4.73  &  0.02    & I/II   \\
s6 (31)  & 3:07:26.45 & 58:30:52.62 &  9.47 &  0.01  &   8.89  & 0.02   &   6.63  &  0.01  &   4.53  &  0.01    & I      \\
s7 (11)  & 3:07:26.50 & 58:30:25.42 &  9.60 &  0.02  &   9.40  & 0.02   &   6.63  &  0.02  &   4.85  &  0.05    & I/II   \\
s8 (36)  & 3:07:26.53 & 58:31:08.32 & 10.76 &  0.03  &  10.24  & 0.02   &   8.98  &  0.02  &   7.32  &  0.04    & I      \\
s9        & 3:07:27.38 & 58:30:12.28 & 11.97 &  0.02  &  11.16  & 0.01   &  10.51  &  0.06  &   9.36  &  0.09    & I      \\
s10 (40)  & 3:07:27.40 & 58:31:15.76 & 10.05 &  0.02  &   9.66  & 0.02   &   8.36  &  0.06  &   6.46  &  0.06    & I/II   \\
s11 (17)  & 3:07:27.76 & 58:30:35.32 & 10.66 &  0.02  &   9.69  & 0.01   &   8.87  &  0.04  &   7.66  &  0.05    & I      \\
s12       & 3:07:28.69 & 58:30:47.39 & 12.68 &  0.03  &  11.22  & 0.01   &  10.81  &  0.08  &  10.30  &  0.22    & I      \\
s13       & 3:07:30.22 & 58:30:58.88 & 13.43 &  0.02  &  12.65  & 0.02   &  12.06  &  0.09  &  10.81  &  0.12    & I      \\
s14       & 3:07:30.83 & 58:31:37.84 & 15.62 &  0.07  &  14.03  & 0.02   &  12.99  &  0.14  &  11.49  &  0.16    & I      \\
s15       & 3:07:31.32 & 58:31:12.45 & 13.20 &  0.01  &  12.53  & 0.01   &  11.74  &  0.03  &  10.35  &  0.03    & I      \\
s16       & 3:07:31.56 & 58:29:59.03 & 11.86 &  0.01  &  11.67  & 0.01   &   9.22  &  0.01  &   7.18  &  0.01    & I/II   \\
s17       & 3:07:32.60 & 58:31:23.03 & 13.94 &  0.02  &  13.72  & 0.02   &  12.70  &  0.13  &  11.11  &  0.22    & I/II   \\

\hline 
\end{tabular}
\end{table*}

\subsection{IRAC Ratio Maps}
As mentioned earlier, the IRAC bands contain a number of prominent molecular lines/features. 
Ch1 contains H$_2$ vibrational-rotational lines while Ch2-4 mostly contain pure rotational lines. 
Ch1, 3 and 4 also contain the PAH features at 3.3, 6.2, 7.7 and 8.6 $\mu$m; 
but Ch2 does not include any PAH features.
\citet {demui90} have detected some of the above mentioned PAH features in AFGL 437. 
As mentioned earlier, \citet{Davis98} have reported some `wisp'-like features of H$_2$ emission (in 2.121 $\mu$m 
narrow band filter that contains the 1-0S(1) line) at a few locations in AFGL 437. 
Several authors have utilised the ratios of IRAC bands to identify some of the molecular 
diagnostics mentioned above (e.g., \citet{Smith05,Povich07,Neufeld08}). 
Since it is difficult to assess the contribution of different molecular transitions to 
different channels, the ratio maps are only indicative; until/unless supplemented by 
spectroscopic evidence. 

The Ch2 is more sensitive to H$_{2}$ lines of high excitation temperatures 
while the Ch4 represents rotational lines of low excitation temperatures (\citet{Neufeld08}).
Likewise, the Ch2 does not have any PAH features while Ch4 has.  
Thus, in the ratio image of Ch2/Ch4, the brighter regions indicate emission regions 
from higher excitations from H$_{2}$ and the darker regions indicate PAH emission. 
This trend is reversed in the image of Ch4/Ch2 (i.e., bright regions show PAH emission and 
dark regions the H$_{2}$ emission). 

Fig 4 shows the ratio maps of Ch2/Ch4 (a) and Ch4/Ch2 (b) in a region surrounding the 
central cluster. The bright regions in Fig 4a probably correspond to H$_{2}$ emission; 
and those in Fig 4b correspond to PAH emission.  
The ratio map Ch2/Ch4 (Fig 4a) brings out well the cross-like features 
towards the south-west edge of the cluster; 
positions of which match well with the faint `wisps' seen in \citet{Davis98}. These
emission features could be the result of the expansion of HII regions 
or stellar winds interacting with the local ISM. 
We have looked for PAH features by examining the ratio images 
of Ch4/Ch2; Ch3/Ch2 and Ch1/Ch2 (since Ch2 does not have any PAH features). 
Fig 4b shows the ratio map of Ch4/Ch2.
The ratio map shows bright regions around AFGL 437W, s5, s7 and s10;
as well as towards north of the NE corner of the boxcar nebula. 
The ratio Ch3/Ch2 (possible indicator for the 6.2 $\mu$m PAH feature) 
does not show any bright features.
The requisite UV photon flux for exciting the PAH features possibly 
comes from the source AFGL 437W. The ratio map shows 
the boxcar/cross shaped feature in dark, probably because of lack of PAH emission. 
The bright narrow linear filament  
seen to the right of AFGL 437W (between the dashed lines in Fig 4b)
indicates probably the ionization front from the massive star 
in the wake of which the PAH is excited (\citet{Povich07}). The detection of the `PAH filament'
corroborates the blister model for AFGL 437W presented by \citet{Wynn81}. 
 
We have examined more closely the ratio map of Ch2/Ch4 to look 
for the infrared counterpart of molecular outflow originating from 
the highly embedded YSO WK34.  
Fig 5 shows a zoomed-in image of Ch2/Ch4 ratio contours over-laid on the Ch4 image.
One can notice the prominent lobe/outflow  
stretched northwards of WK34 and slightly bent towards NE, with a total extension of $\sim$ 0.16 pc.  
The outflow direction and size are consistent with the earlier reports. 
The bending itself is attributed to the presence of the nearby source AFGL 437N (\citet{Wein96}).  

\begin{table*}
\centering
\caption{Physical Parameters derived from SED modelling of the main sources AFGL 437 W,S,N and WK34 (see text)}
\label{tab2}
\begin{tabular}{lcccccccc}
    \hline 
Source  & Age           & $M_*$         &     $T_*$       & $L_*$           &  $\dot{M}_{env}$           & $\dot{M}_{disk}$	        & $A_v$ &Degeneracy/   \\
Name    & log(yr)       & M$_{\odot}$   &log(T(K))   & log(L$_{\odot}$)& log(M$_{\odot}$ yr$^{-1}$) & log(M$_{\odot}$ yr$^{-1}$)   & mag   &  No. of models \\
\hline 
W       & 5.36$\pm$0.19 & 8.28$\pm$1.48 & 4.28$\pm$0.11   & 3.55$\pm$0.23   & -4.84$\pm$0.17       & -7.01$\pm$1.16             &  4.0$\pm$0.7&    16    \\    
S       & 4.35$\pm$0.90 & 7.79$\pm$1.72 & 3.86$\pm$0.31   & 3.14$\pm$0.39   & -4.12$\pm$1.30       & -5.84$\pm$1.12             &  9.2$\pm$1.8 &    72     \\	 
N       & 4.82$\pm$0.46 & 9.36$\pm$1.62 & 3.99$\pm$0.19   & 3.48$\pm$0.30   & -3.96$\pm$0.94       & -6.02$\pm$0.84             & 18.0$\pm$8.0  &    132     \\	  
WK34    & 3.94$\pm$0.57 & 7.23$\pm$1.91 & 3.71$\pm$0.10   & 3.00$\pm$0.30   & -4.06$\pm$0.55       & -4.69$\pm$0.39             & 23.0$\pm$5.6  &	3      \\	   
\hline 
\end{tabular}
\end{table*}

\subsection{SED modelling of Cluster Sources} 
 By modelling the SEDs of the identified YSOs, we derive physical parameters 
of both their photospheres and mass accreting disks/envelopes. 
For this purpose, we constructed the SEDs for all the 21 sources listed in Table 1, 
from the optical to sub-mm data depending upon the availability in the archives and 
published literature. 
The optical data\footnote[2]{from  
http://vizier.u-strasbg.fr/viz-bin/VizieR} 
are available only for the sources W (B,V bands), s7 (V upper limit) 
and s16 (B,V,R upper limits). In fact, the lack of optical data is indicative of the large extinction that  
most of the sources suffer. 
The JHK fluxes for the sources s10, s16 and s17 are taken from 2MASS  
and for s2, s3, s5-s8 and s11 from \citet{Wein96}.  
While the source s1 has only K band flux from \citet{Wein96}, 
s4, s9, and s12-s15 do not have near-infrared counterparts. 
The JHK photometric fluxes for WK 34 are taken from \cite{Meakin05}.
We have used sub-mm/mm data (\citet{Dent98}) as upper limits for WK34, S, W and N
(because of the large beam sizes that nearly encompass the entire central region of the cluster), 
only to constrain the models.
The rest of the sources listed in Table 1 are outside the field of view of 
the sub-mm observations. 
The SEDs are then modelled  
using an on-line 2D-radiative transfer  
tool due of \citet{Robit06}, which assumes an accretion scenario with a central 
star surrounded by an accretion disk, an in-falling 
flattened envelope and bipolar cavities. This tool was successfully tested by \citet{Robit07}, 
on a sample of low mass YSOs and  by \citet{Grave09} on high mass protostars. 
The SED model tool requires a minimum of three data points that 
are of good quality. 
In our sample all the 21 sources have at least 
three such data points. For a source that has meager number of data points,  
clearly the tool picks out a large number of solutions 
that can fit the data well, within the specified limit on $\chi^{2}$. 
If the SED has larger data, spread over the  
wavelength region of 0.5 to 1000 $\mu$m, then the model would be better constrained 
to yield results with least standard deviations \citep{Robit07}. 
The distance to the source and visual extinction are to be  
given as input parameters, usually as a range of values. This leads to 
a further degeneracy of the models. 
In our case, however, the cluster distance is fairly well-determined 
and the extinctions were estimated from the available JHK data on individual sources. 
But, in order to avoid `over-interpretation' of SEDs \citep{Robit07}, we have provided a range of visual extinction values 
for each object to account for the uncertainties in their determination from JHK photometry. 
Only those models are accepted which satisfy the criterion of $\chi^{2}$ - $\chi^{2}_{best}$ $<$ 3. 

We show in Fig~\ref{fig6} the SED modelling results for the sources AFGL 437W, S, N and WK34. 
Table~\ref{tab2} gives the weighted mean values of the derived physical parameters for the four sources 
along with the standard deviations. The table also lists the model-derived weighted mean values of A$_{v}$  
with standard deviations and the 
degeneracy of the models (i.e., the number of solutions that satisfy the criterion mentioned above).
The model-derived parameters listed in Table 2 indicate that all the four sources are likely to be massive (early B type). 
This is consistent (within the standard deviations) with the observations on W and S that are associated with UCHII regions.
It may be noted that observationally very little is known about the spectral type of the source N.

The model parameters for WK34 suggest that the source is massive but of young age with 
effective temperature still not sufficient to create an HII region (see Table 2). Its low luminosity 
is suggested by earlier workers also. This is also reflected in its  
outflow, which shows emission in H$_{2}$ (as revealed by the ratio map in Fig 5) but not in PAH;  
indicating that WK34 is still not hot enough to produce sufficient UV flux. 
The modelling suggests that the source AFGL 437W and S are also massive stars but more evolved to attain sufficiently 
large effective temperature to excite an HII region. 
\citet{Qin08} derived a CO outflow entrainment
rate of 7.4$\times10^{-4}$ M$_{\odot}$/yr for the outflow source (WK34), which agrees with that observed  
for massive stars (e.g., \citet{Beuther02}). In comparison, the low mass stars 
show much less entrainment rates (\citet{Wu96}). In a study of the molecular outflows from high mass 
YSOs, \citet{Ridg01} have concluded that these outflows are often poorly collimated, which again 
points to the possibility that WK34 is a massive protostellar object.    

From the SED modelling of the 21 sources, we infer that 
the weighted mean values of the masses and ages of the YSOs in the central cluster 
(for the 21 sources listed in Table 1) are 
in the ranges 1 - 10 M$_{\odot}$ and 10$^{4.1 - 6.4}$ yr respectively; while 
the luminosities are in the range of 10$^{0.47-3.48}$ L$_{\odot}$.

\section{Conclusions}
The important conclusions of this work are as follows:
\begin{enumerate}
\item {\it Spitzer} IRAC imaging photometry is presented on the massive star forming region AFGL 437;
\item Several new embedded YSOs are identified within 90 arcsec of the central compact cluster;
\item The IRAC ratio maps indicate molecular outflow corresponding to WK34 which is possibly due to H$_2$; 
\item The SED modelling of the outflow driving source WK34 indicates that 
it is a massive but very young protostar not yet able to drive a HII region;
\item SED modelling of the 21 sources gives their masses in the range 1-10 M$_{\odot}$, ages 10$^{4.1 - 6.4}$ yr 
and luminosities 10$^{0.47-3.48}$ L$_{\odot}$. 
\end{enumerate}

\begin{figure*}
\includegraphics[width=\textwidth]{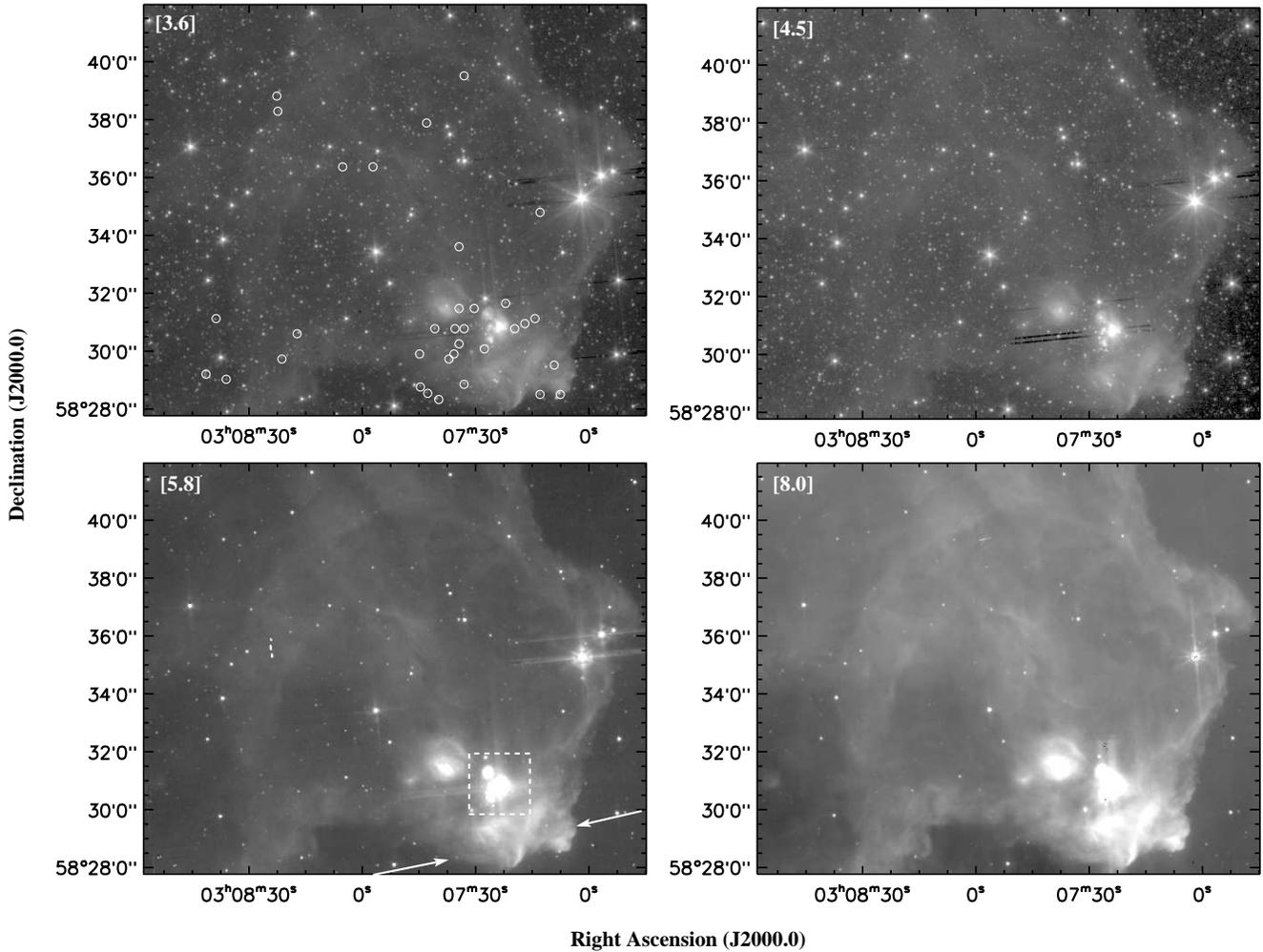}
\caption {IRAC images of AFGL 437 (size $\sim$ 17.4 $\times$ 14.2 arcmin$^2$) are shown in all four channels 
(in log scale). The central compact cluster is marked by the square box (of side $\sim$ 129 arcsec or 1.2 pc) 
in Ch3 image (bottom left). The associated bubble-like 
extended nebulosity is of size $\sim$ 8.0 pc in the SW-NE direction. The cross-like or boxcar-shaped 
dense nebula expanding into the local ISM is shown marked by the two arrows in the Ch3 image (bottom left).
The open circles in Ch1 (top left) mark the Class II sources identified from IRAC photometry.}
\label{fig1}
\end{figure*}

\begin{figure*}
\includegraphics[width=14cm]{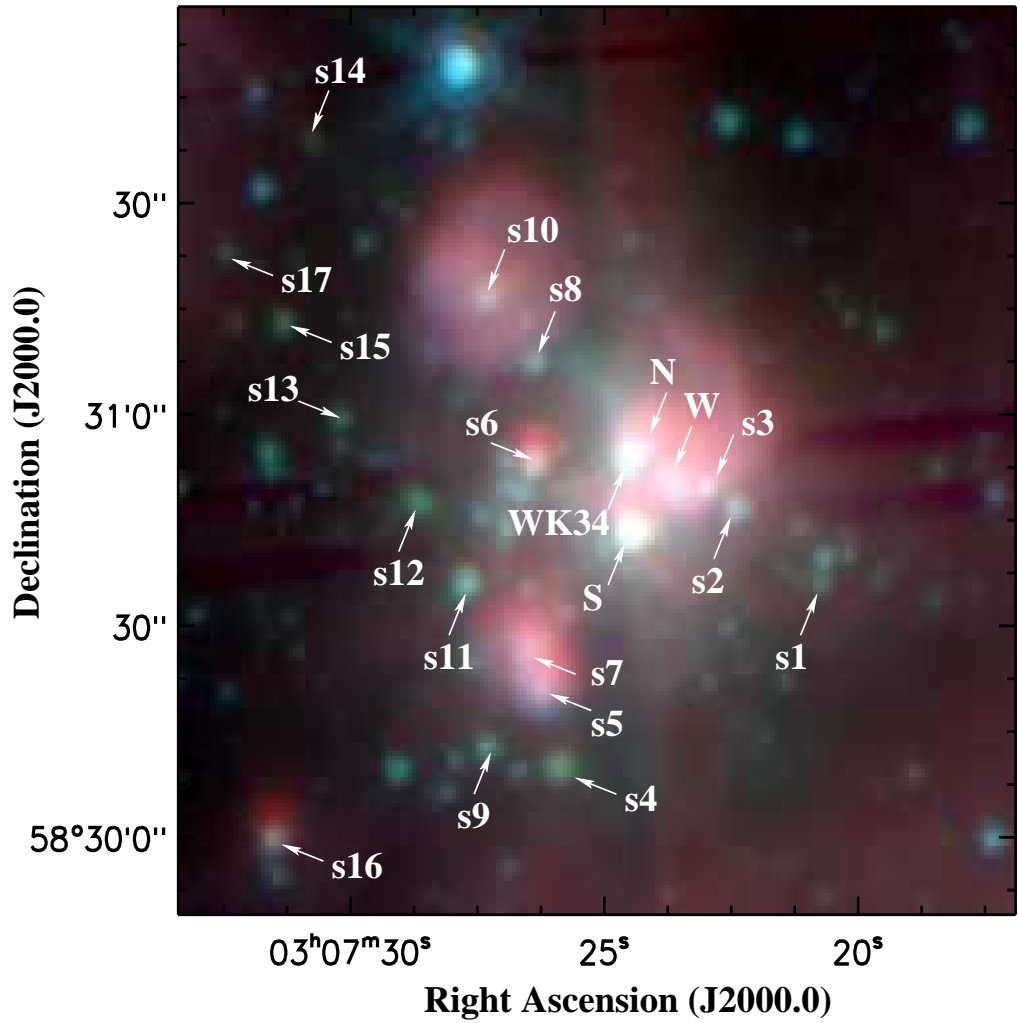}
\caption{The compact cluster (see the box in Fig 1) is shown in a colour composite zoomed-in 
image (8.5 (red), 4.5 (green) and 3.6 (blue) $\mu$m). 
The identified YSOs are shown as s1, s2.. along with the brighter sources AFGL 437S, W and WK34 (see Table 1).}
\label{fig2}
\end{figure*}

\begin{figure*}
\includegraphics[width=14cm]{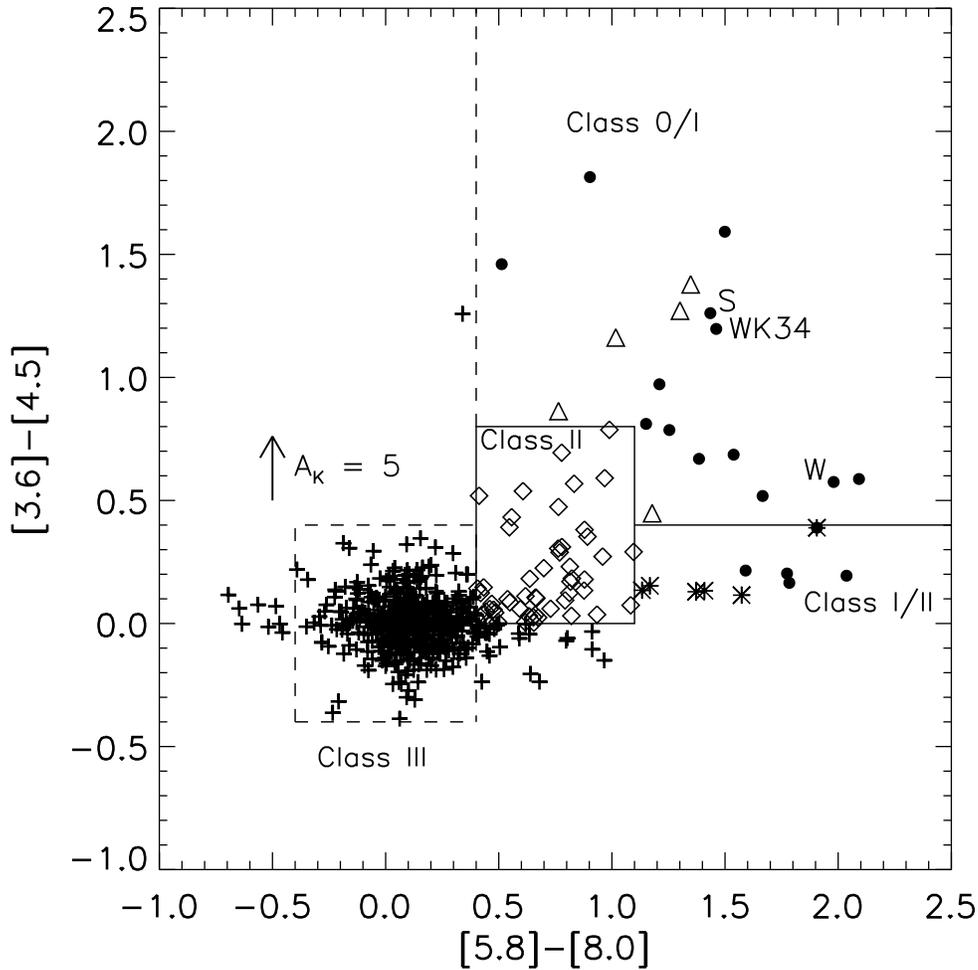}
\caption{Mid-IR colour-colour diagram constructed using the Spitzer IRAC bands. The boxes indicate possible regions of 
Class 0/I, Class II, and Class III sources based on \citep{Allen04}. Class I/II sources share the properties of both 
Class I and Class II sources. The filled circles represent the sources within 64 arcsec of the centre of the cluster
(see Table 1 and Fig 2); the asterisks are Class I/II stars and the diamonds are for Class II (see Fig 1 for their locations). 
The Class III stars are shown as plus-signs. The triangles are Class I sources found quite far away from 
the central cluster. The principal sources of the cluster, AFGL 437 S, W, and WK34 
can be seen marked 
inside the box designated for Class 0/I protostars. Extinction vector for K band, 
A$_{K}$ =5 is shown in the diagram, which is 
calculated using averaged reddening law from \citet{Flaherty07}.} 
\label{fig3}
\end{figure*}

\begin{figure*}
 \includegraphics[width=\textwidth]{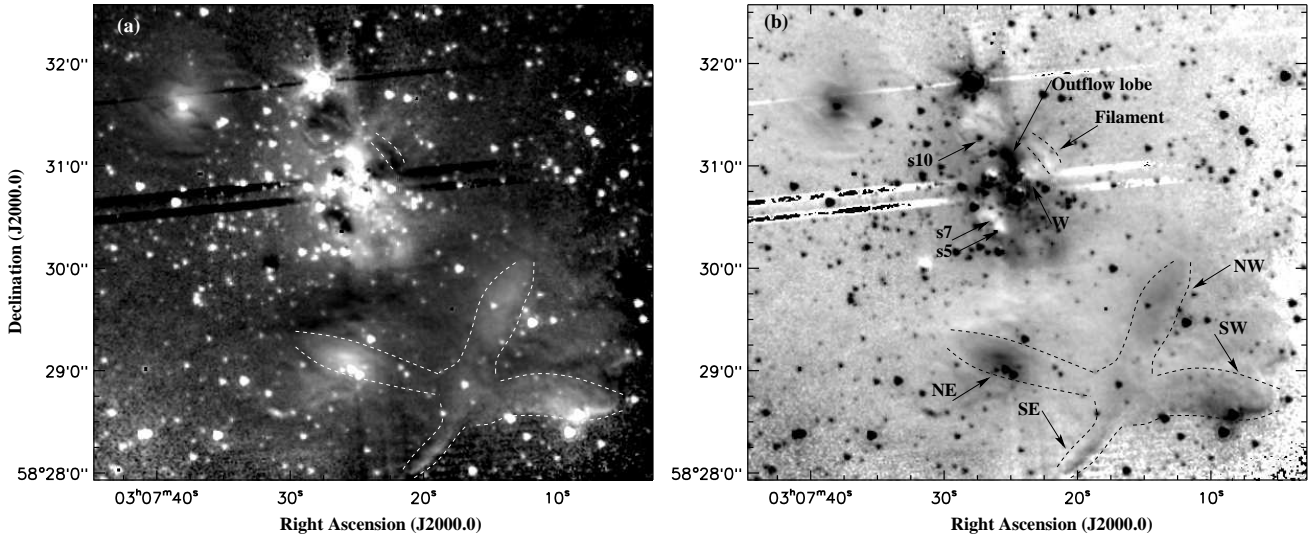}
 \caption{IRAC Ch2/Ch4 (a) and Ch4/Ch2 (b) ratio images in log scale (from mosaics made with pixel ratio of 2). 
 The central cluster and the 
 cross- or boxcar-like structure to its south-west can be noticed. This structure has a size of 
 1.7 pc in the NE-SW and 1.5 pc in NW-SE directions. The bright patches in the image on the left (a) 
 show regions dominated by H$_{2}$ emission; while in the image on the right (b) the bright patches 
 indicate regions that are emitting PAH bands. The prominent outflow lobe, some of the YSOs, 
 and the H$_2$ emission patches in the four corners of the cross-like structure are marked (by dashed curves).
 The near-horizontal tracks across the image are artifacts. The filament-like structure to the right of AFGL 437W 
 is marked by dashed lines in the upper half of the figure.} 
 \label{fig4}
 \end{figure*}

\begin{figure*}
\includegraphics[width=14cm]{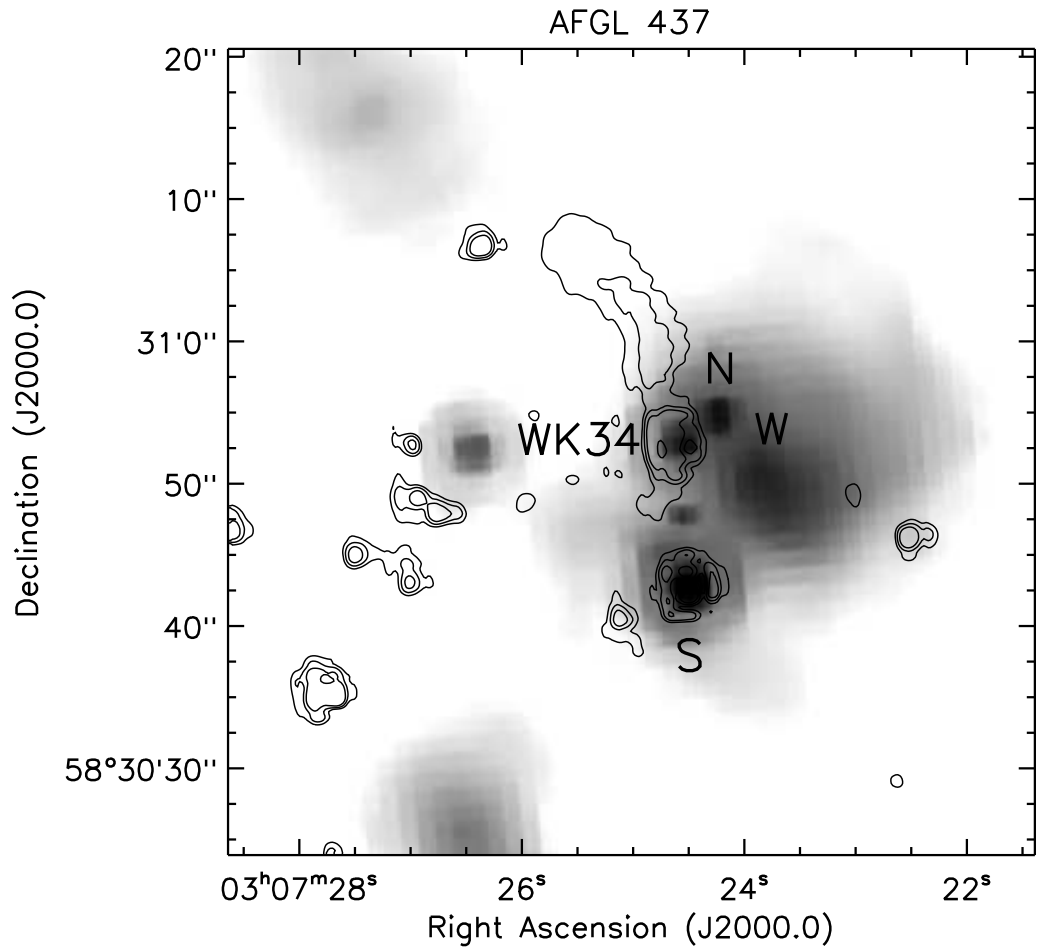}
\caption{IRAC 8 $\mu$m zoomed-in image of the cluster is shown in log scale overlaid by the contours of IRAC ratio 
image of (Ch2/Ch4). Contours show the outflow lobes associated with the source WK34.
Outflow lobes are stretched in north direction with a bend towards east. 
The IRAC Ch 4 (8 $\mu$m) image is made with a pixel ratio of 8 in a spatial extent of 
$\sim$ 55 $\times$ 55 arcsec$^2$; while the ratio contours are plotted 
with a minimum of 0.178 and maximum of 8 MJy/Sr.} 
\label{fig5}
\end{figure*}

\begin{figure*}
\includegraphics[width=\textwidth]{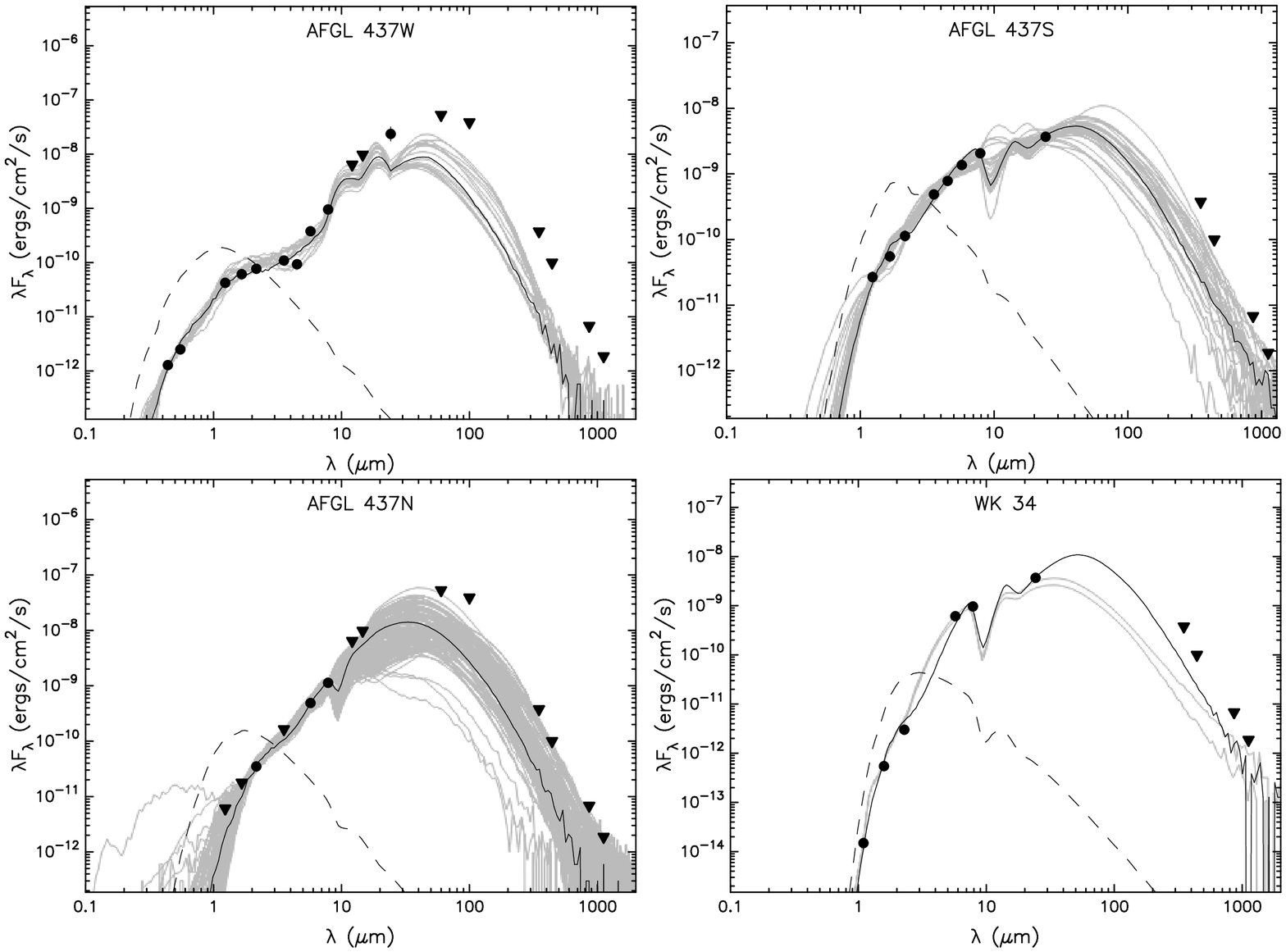}
\caption{Spectral Energy Distributions for four YSOs in the central cluster: AFGL 437W, S (top panels) 
N and WK34 (bottom panels). 
Filled circles are observed fluxes of good quality (with filled triangles as upper limits) 
taken from archives or published literature (see text for references) 
and the curves show the model fits. The thin black curve corresponds to the best model.  
The dashed curves represent photospheric contributions. 
The model parameters of the YSOs are listed in Table 2.} 
\label{fig6}
\end{figure*}

\section*{Acknowledgments}

The research work at PRL is supported by the Department of Space, Government of India. 
This work is based (in part) on observations made with the 
Spitzer Space Telescope, which is operated by the Jet Propulsion Laboratory, 
California Institute of Technology under a contract with NASA.
We acknowledge the use of data from the 2MASS, 
which is a joint project of the University of Massachusetts and the 
Infrared Processing and Analysis Center/California Institute of Technology, 
funded by the NASA and the NSF. We thank the anonymous referee for constructive 
comments.

\label{lastpage}


\begin{thebibliography}{99}

\bibitem[\protect\citeauthoryear{Allen et al.}{2004}]{Allen04}
Allen L.~E. et al., 2004, ApJS, 154, 363

\bibitem[\protect\citeauthoryear{Arquilla \& Goldsmith}{1984}]{Arquilla84}
Arquilla R., and Goldsmith P.~F., 1984, ApJ, 279, 664

\bibitem[\protect\citeauthoryear{Beuther et al.}{2002}]{Beuther02}
Beuther H., Schilke P., Sridharan T.~K., Meuten K.~M., Walmsley C.~M., Wyrowski F., 2002, A\&A, 383, 892

\bibitem[\protect\citeauthoryear{Cohen \& Kuhi}{1977}]{Cohen77}
Cohen M., and  Kuhi L.~V., 1977, PASP,89,829

\bibitem[\protect\citeauthoryear{Davis et al.}{1998}]{Davis98}
Davis C.~J.,G. Moriarty-schieven G., Eisloffel J., Hoare M.~G., and Ray T.~P., 1998, AJ,115,1118

\bibitem[\protect\citeauthoryear{de Muizon et al.}{1990}]{demui90}
de Muizon Jourdain M., d'Hendecourt L., Geballe T.~R., 1990, A\&A, 227, 526

\bibitem[\protect\citeauthoryear{de Wit et al.}{2009}]{deWit09}  
 de Wit W.~J. et al., 2009, A\&A, 494, 157 

\bibitem[\protect\citeauthoryear{Dent et al.}{1998}]{Dent98}
Dent W.~R.~F., Matthews H.~E., and Ward-Thompson D., 1998, MNRAS, 301,1049

\bibitem[\protect\citeauthoryear{Devine et al.}{2008}]{Devine08}
Devine, K.E., Churchwell, E.B., Indebetouw, R., Watson, C., and Crawford, S.M., 2008, AJ, 135, 2095 

\bibitem[\protect\citeauthoryear{Fazio et al.}{2004}]{Fazio04}
Fazio G.~G. et al., 2004, ApJS, 154, 10  

\bibitem[\protect\citeauthoryear{Flaherty et al.}{2007}]{Flaherty07}
Flaherty K.~M., Pipher J.~L., Megeath S.~T., Winston E.~M., Gutermuth R.~A., Muzerolle J., 
Allen, L.~E., and Fazio, G.~G., 2007, ApJ, 663, 1069

\bibitem[\protect\citeauthoryear{Gomez et al.}{1992}]{Gomez92}
Gomez J.~F., Torelles J.~M., Estalalella R., Anglada G., Vades-Montenegro L., and Ho P.~T.~P., 1992, ApJ, 397, 492 

 \bibitem[\protect\citeauthoryear{Grave \& Kumar}{2009}]{Grave09}
Grave J.~M.~C. and Kumar M.~S.~N., 2009, A\&A, 498, 147

\bibitem[\protect\citeauthoryear{Kleinmann et al.}{1977}]{Kleinmann77}
Kleinmann S.~G., Sargent D.~G., Gillet F.~C., Grasdalen G.~L., and Joyce R.~R., 1977, ApJ (Letters), 215, L79

\bibitem[\protect\citeauthoryear{Kurtz, Churchwell \& Wood}{1994}]{Kurtz94}
 Kurtz S., Churchwell E., and Wood D.~O.~S., 1994, ApJS, 91, 659

\bibitem[\protect\citeauthoryear{Lada \& Lada}{2003}]{Lada03}
 Lada, C.J., and Lada, E.A., 2003, ARA\&A, 41, 57

\bibitem[\protect\citeauthoryear{Meakin et al.}{2005}]{Meakin05}
Meakin C.~A., Hines D.~C., and Thompson R.~I., 2005, ApJ, 634, 1146

\bibitem[\protect\citeauthoryear{Megeath et al.}{2004}]{Megeath04} 
Megeath S.~T. et al., 2004, ApJS, 154, 367 

\bibitem[\protect\citeauthoryear{Neufeld \& Yuan }{2008}]{Neufeld08}
Neufeld D.~A., Yuan Y., 2008, ApJ, 678, 974

\bibitem[\protect\citeauthoryear{Povich et al.}{2007}]{Povich07} 
Povich M.~S. et al., 2007, ApJ, 660, 346

\bibitem[\protect\citeauthoryear{Qin et al.}{2008}]{Qin08}
Qin, S.-L., Wang, J.-J., Zhao, G., Miller, M., and Zhao, J.-H., 2008, A\&A, 484, 361 
 
\bibitem[\protect\citeauthoryear{Rainer \& McLean}{1987}]{Rainer87}
Rainer J., and McLean I., 1987, in Infrared Astronomy with Arrays, ed. C. G.
Wynn-Williams, E. E. Becklin \& L. H. Good (Honolulu: Univ. of Hawaii), 272 

\bibitem[\protect\citeauthoryear{Ridge \& Moore}{2001}]{Ridg01}
Ridge, N.A., \& Moore, T.J.T., 2001, A\&A, 378, 495

 \bibitem[\protect\citeauthoryear{Robitaille et al.}{2006}]{Robit06} 
Robitaille T.~P., Whitney B.~A., Indebetouw R., Wood K., and Denzmore P., 2006, ApJS, 167, 256

 \bibitem[\protect\citeauthoryear{Robitaille et al.}{2007}]{Robit07} 
Robitaille T.~P., Whitney B.~A., Indebetouw R., and Wood K., 2007, ApJS, 169, 328

 \bibitem[\protect\citeauthoryear{Saito et al.}{2006}]{Saito06} 
Saito H., Saito M., Moriguchi Y., and Fukui Y., 2006, PASJ, 58, 343

 \bibitem[\protect\citeauthoryear{Saito et al.}{2007}]{Saito07} 
Saito H., Saito M., Sunada K., and Yonekura Y., 2007, ApJ, 659, 459

 \bibitem[\protect\citeauthoryear{Skrutskie et al.}{2006}]{Skru06}
Skrutskie, M.~F. et al., 2006, AJ, 131, 1163

\bibitem[\protect\citeauthoryear{Smith \& Rosen}{2005}]{Smith05}
Smith, M.~D., and Rosen, A., 2005, MNRAS, 357, 1370 

\bibitem[\protect\citeauthoryear{Torrelles et al.}{1992}]{Torrelles92}
Torrelles J.~M., Gomez J.~F., Anglada G., Estalella R., Mauersberger R., and Eiroa C., 1992, ApJ, 392, 616

\bibitem[\protect\citeauthoryear{Weintraub et al.}{1996}]{Weinet96} 
 Weintraub D.~A., Kastner J.~H., Gatley I., and  Merrill K.~M., 1996, ApJ, 468, L45 

\bibitem[\protect\citeauthoryear{Weintraub \& Kastner}{1996}]{Wein96} 
 Weintraub D.~A., and Kastner J.~H., 1996, ApJ, 458, 670

\bibitem[\protect\citeauthoryear{Wu et al.}{1996}]{Wu96}
Wu, Y., Huang, M., and He, J., 1996, A\&AS, 115, 283W 

\bibitem[\protect\citeauthoryear{Wynn-Williams et al.}{1981}]{Wynn81}
Wynn-Williams C.~G., Becklin E.~E., Beichman C.~A., Capps R., Shakeshaft J.~R., 1981, ApJ, 246, 801

\end{thebibliography}
\end{document}